\documentclass[11pt,onecolumn,amssymb,nofootinbib]{revtex4}
\usepackage{amsmath, amsthm, amscd, amssymb}
\usepackage{graphicx, braket}
\usepackage{bm}
\usepackage{bbm}
\usepackage{epstopdf}

\begin{document}

\title{\bf Trace dynamics and its implications for my work of the last two decades} \bigskip

\author{Stephen L. Adler}
\email{adler@ias.edu} \affiliation{Institute for Advanced Study,
Einstein Drive, Princeton, NJ 08540, USA.}

\begin{abstract}
I review the basic ideas of ``trace dynamics'', as formulated in my 2004 Cambridge University Press book ``Quantum Theory as an Emergent Phenomenon'', and then discuss how they have influenced much of my work of the last two decades.

\end{abstract}

\maketitle
\section{Trace dynamics as an underlying dynamics for quantum field theory}

``Trace Dynamics''  \cite{td} grew out of a paper entitled ``Generalized Quantum Dynamcs''  \cite{gqd} that I wrote  with my then graduate student Andrew Millard.  The basic idea was to construct an embedding of standard quantum mechanics in a larger mathematical framework, with the aim of understanding the strange nonlocal aspects of entanglement in quantum theory, as manifested in particular in the quantum measurement problem. In earlier work I had studied the extension of standard complex quantum mechanics  to quaternionic quantum mechanics \cite{qqm},  which motivated  the trace variational  principle that was later used in trace dynamics. But quaternionic quantum mechanics  is still linear, and suffers from the same measurement paradoxes as its complex cousin, motivating the search for a manifestly nonlocal extension of the standard quantum formalism.

Trace dynamics is a Lagrangian and Hamiltonian theory of completely non-commuting variables $q$'s (generalized coordinates) and $p$'s (generalized momenta), which can be bosonic (Grassmann even) or fermionic (Grassmann odd) matrices or operators. Using a dot to denote the time derivative, the trace Lagrangian ${\bf L}$ is formed as a trace over a polynomial $L$ in the $q$'s and $\dot{q}$'s, and operator ordering issues are resolved by assuming validity of cyclic permutation under the trace.  So we have as  our starting point
\begin{equation}\label{traceL}
{\bf L}={\rm Tr} L[\{q_r\},\{p_r\}]~~~,
\end{equation}
with the subscript $r$ denoting degrees of freedom.

Now form the variation $\delta {\bf L}$ and cyclically reorder factors in each term of $L$ so that all $\delta q_r$ and $\delta \dot q_r$ stand on the right.  This motivates the definition of derivatives of $\bf L$,
\begin{equation}\label{deriv}
\delta {\bf L}={\rm Tr} \sum_r \left(\frac{\delta{\bf L}}{\delta q_r} \delta q_r +\frac{\delta{\bf L}}{\delta \dot q_r} \delta \dot q_r\right)~~~.
\end{equation}

One can now show that defining a trace action ${\bf S}=\int_{-\infty}^{\infty} dt {\bf L}$,   the stationary action principle $\delta {\bf S}=0$ directly implies operator Euler-Lagrange equations
\begin{equation}\label{euler}
\frac{\delta{\bf L}}{\delta q_r} -\frac {d}{dt} \frac{\delta{\bf L}}{\delta \dot q_r} =0~~~,
\end{equation}
without invoking a canonical quantization procedure.  Thus trace dynamics gives a dynamics that is more general than standard quantum mechanics, but which can be shown \cite{td} to reduce to standard quantum theory when restricted to $q$'s and $p$'s that obey the Heisenberg algebra.

To recover quantum theory in the general case of non-commuting variables, we consider the equilibrium statistical mechanics of trace dynamics.  There are three generic conserved quantities:

\begin{enumerate}
\item {\bf Trace Hamiltonian}
Let us define
\begin{align}\label{pandham}
p_r=& \frac{\delta{\bf L}}{\delta \dot q_r}~~~,\cr
{\bf H}=& {\rm Tr}\sum_r p_r \dot q_r - {\bf L}~~~.\cr
\end{align}
Then taking taking variations we have
\begin{equation}\label{hamvar}
\frac{\delta{\bf H}}{\delta q_r}=-\dot p_r~~~,~~~ \frac{\delta{\bf H}}{\delta  p_r} =\dot q_r~~~,
\end{equation}
and using the Euler-Lagrange equations we see that the time derivative of $\bf H$ vanishes,
\begin{align}\label{Hcons}
\frac{\bf H}{dt}=&{\rm Tr} \sum_r\left( \frac{\delta{\bf H}}{\delta q_r} \dot q_r +  \frac{\delta{\bf H}}{\delta p_r}  \dot p_r \right) \cr
=& {\rm Tr} \sum_r \left(  -\dot p_r \dot q_r + \dot q_r \dot p_r \right) =0 ~~~.\cr
\end{align}

\item{\bf Operator $\tilde C$}
The operator  $\tilde C$ is defined by
\begin{equation}\label{tidleC}
\tilde C=\sum_{r, {\rm bosonic}}[q_r,p_r]- \sum_{r,{\rm fermionic}}\{q_r,p_r\}~~~,
\end{equation}
where $[,]$ denotes the commutator and $\{,\}$ denotes the anticommutator.  When the trace Hamiltonian is global unitary invariant, then $\tilde C$ is the corresponding Noether charge operator, and is conserved,
\begin{equation}\label{ccons}
\frac{d \tilde C}{dt}=0~~~.
\end{equation}

\item{\bf Phase space measure}  Let us define the
phase space measure $d \mu$ by
\begin{equation}\label{mudef}
d\mu=\prod_{r,m,n,A} d\langle m|q_r|n \rangle^A d\langle m|p_r|n\rangle^A~~~,
\end{equation}
where $r$ as before labels the matrix degrees of freedom, $m,n$ are matrix element labels, and $A=0,1$ labels the real and imaginary parts of the complex number-valued matrix element.  Then one can show that $d\mu$ is invariant under the canonical transformation with generator $\bf G$
\begin{equation}\label{canon}
\delta p_r= -\frac{\delta {\bf G}}{\delta q_r} ~~~,~~ \delta q_r= \frac{\delta {\bf G}}{\delta p_r}~~~.
\end{equation}
Hamiltonian evolution is a special case of this canonical transformation, so $d\mu$ is conserved, giving a generalized Liouville theorem, and implying that we can use the methods of canonical ensemble statistical mechanics.

\end{enumerate}

The canonical ensemble $\rho$ in trace dynamics is defined by
\begin{equation}\label{canens}
\rho=Z^{-1} e^{-\tau {\bf H} - {\rm Tr} \tilde \lambda \tilde C}~~~,
\end{equation}
where the partition function $Z$ is given by
\begin{equation}\label{partit}
Z=\int d\mu   e^{-\tau {\bf H} - {\rm Tr} \tilde \lambda \tilde C}~~~,
\end{equation}
which guarantees that  $\int d\mu \rho=1$.

We can now form averages over the canonical ensemble of polynomials in the degrees of freedom.  Of particular note, we write the average of $\tilde C$ in the standard form
\begin{align}\label{avC}
\langle \tilde C  \rangle_{\rm AV}=&i_{\rm eff} D~~~,\cr
i_{\rm eff}^2=&-1~~,~~~ i_{\rm eff}^\dagger = -i_{\rm eff}~~,~~~[i_{\rm eff}, D]=0~~~.\cr
\end{align}
In the simplest case, which we assume, $D$ is a multiple of the unit matrix, so that
\begin{equation}\label{hbar}
D=\hbar 1~~~,
\end{equation}
where $\hbar$ is a constant with dimensions of action. Since $\tilde C$ is traceless, then so is $i_{\rm eff}$,
\begin{equation}\label{trieff}
{\rm Tr}i_{\rm eff}=0~~~.
\end{equation}

We have now given enough equations to convey the basic idea of trace dynamics.  Using the canonical ensemble, and taking averages over this ensemble, one finds that equipartition arguments give an emergent structure with the form of quantum field theory, provided the underlying trace dynamics theory  develops a scale hierarchy in which variations of the trace Hamiltonian decouple, and provided that the numbers of bosonic and fermionic degrees of freedom are essentially equal.  In this theory $i_{\rm eff}$, defined through the ensemble average of the conserved Noether charge $\tilde C$, plays the role of the complex unit $i$ of standard complex quantum field theory. For further details see the book \cite{td}, and the YouTube recording of my OSMU lecture \cite{osmu}, which also gives further detailed formulas.

\section{How trace dynamics ideas have influenced my work of the last two decades}

A large part of my work since 2000 has been related to quantum foundations, trace dynamics, and spinoffs from trace dynamics. Some exceptions have been  (i) a revisit of saturation of current algebra sum rules with Paco Ynduráin \cite{paco}, and a revisit for Scholarpedia of local current algebra sum rules \cite{scholar}, (ii) papers on the so-called ``flyby anomaly'' \cite{flyby},  which in the end was not confirmed by the Juno flyby; however this project led to a strong bound on the mass of dark matter gravitationally bound to Earth below the moon's orbit \cite{moon}, and (iii) my work on multi-dimensional numerical integration methods, summarized in my ``PAMIR'' book \cite{pamir}.  I won't say more here about these projects and some others that are not offshoots of \cite{td}.   Instead, in what follows I briefly discuss the trace-dynamics related projects and give key citations for each.

\begin{itemize}

\item{\bf Connections with the Ghirardi-Rimini-Weber-Pearle  (GRWP) \cite{grw} theory of objective reduction}
According to \cite{td}, quantum field theory emerges as the thermodynamics of an underlying trace dynamics. Thermodynamics always has fluctuation corrections, such as Brownian motion, so it is natural to consider the idea that GRWP represents the effective quantum mechanical realization of Brownian motion corrections to trace dynamics.  I wrote many papers motivated by this idea; a few highlights are:   With Larry Horwitz, I wrote a paper \cite{larry}  completing and simplifying Lane Hughston's compact proof of the Born Rule within the framework of GRWP theory. A paper with Dorje Brody, Todd Brun, and Lane Hughston \cite{dorje} extended this proof to give the L\"uders rule in the case of degeneracies.   In the paper \cite{lower} I gave estimates of lower bounds on the noise coupling in reduction models, based on the hypothesis that formation of virtual images constitutes a measurement.  This gives coupling values suggesting effects in the mesoscopic range, now under experimental test.  With Fethi Ramazano\u glu \cite{fethi}, we worked out the generalization of a prior calculation of photon emission from free electrons driven by this noise, to the case of emission from general atomic and condensed matter systems.  With Angelo Bassi \cite{bassi1}, \cite{bassi2}, we worked out the detailed extension of objective reduction models to non-white noise.  There were  other papers I wrote on measurement theory issues, one of which, my paper with Andrea Vinante on bulk heating effects from collapse models \cite{vinante},
benefited from my early career involvement with condensed matter physics.  A few others were written in a more philosophical vein, such as my paper arguing that decoherence does not solve the measurement problem \cite{decoh}, and a paper on how collapse models can evade the Conway-Kochen ``Free Will Theorem'' \cite{freewill}.

\item{\bf Dark matter as a decoupled $i \to -i$ sector}

In the complex quantum theory emergent from trace dynamics, the role of the imaginary unit is played by $i_{\rm eff}$.  Since $i_{\rm eff}$ has square -1, and is anti-self-adjoint and traceless, when diagonalized it must have equal numbers of eigenvalues $i$ and $-i$.  So the emergent quantum theory has two sectors, which could be weakly coupled, one with imaginary unit $i$ and one with imaginary unit $-i$.  On page 191 of \cite{td} I suggested that the $-i$ sector could be relevant for the dark sector of the universe needed in $\Lambda$CDM cosmology.  In a later Gravitational Essay \cite{essay}, I more specifically suggested that the $-i$ sector could be the source of astrophysical ``dark matter'', with different properties from the ordinary matter in the $i$ sector.  Since the nature of dark matter is still unknown, I plan in the future to try to further develop this idea.

\item{\bf  Unification models for strong and electroweak forces, with boson-fermion balance but no full supersymmetry}

A consistency requirement for trace dynamics to give an emergent quantum theory obeying the Heisenberg algebra is boson-fermion balance, that is, the number of bosons and fermions should be equal or approximately so. This is a weaker requirement than full supersymmetry, suggesting that one might try to formulate grand unification models obeying boson-fermion balance without requiring particles to lie in supermultiplets.  In investigating this, I focused on models with an underlying $SU(8)$ gauge symmetry, containing spin-3/2 fermions as well as spin-1/2 fermions \cite{family}, a scheme motivated in part by a rearrangement of the particle content of $N=8$ supergravity.  This led to studies of gauged Rarita-Schwinger fields \cite{rs1}, \cite{rs2} and symmetry breaking mechanisms for $SU(8)$ based on a third rank antisymmetric tensor scalar field \cite{sym1}, \cite{sym2}, \cite{sym3}.  To get a calculable perturbative spin-3/2 anomaly, a direct coupling of a spin-1/2 field to the spin-3/2 field was introduced, and the properties of this modified spin-3/2 model were analyzed in a number of papers including \cite{ex1}, \cite {ex2}, \cite{ex3}, \cite{ex4}.  The direct coupling changes the spin-3/2 anomaly value from that traditionally reported, requiring the introduction of additional boson fields into my original model \cite{final}.  To get the standard model three families from the model requires an assumption of quark-lepton compositeness, for which there is no current evidence.  I eventually set this investigation aside pending further experimental clues as to what physics beyond the standard model might look like (and so far there have been none).

\item{\bf  Incorporating gravity into trace dynamics, and a Weyl scaling invariant cosmological constant action}

Towards the end of my PAMIR multiple integration project, my wife Sarah got tired of listening to me fuming over debugging issues, and told me it was time for me to start something else.  This was a very useful wakeup call, and got me thinking about an issue that I had neglected before, the incorporation of gravity into trace dynamics.  I pursued this with a classical gravity background metric, and noted that the when the fundamental field degrees of freedom are massless, as I was assuming, the canonical ensemble in trace dynamics is invariant  under the Weyl scaling (the original ``gauge invariance'')
\begin{equation}\label{weylscal}
g_{\mu\nu}(x) \to \lambda^2(x) g_{\mu\nu}(x)~~~,
\end{equation}
with $\lambda(x)$ a general scalar field.  This suggests that the non-derivative part of the induced gravitational action is also Weyl scaling invariant, giving a novel alternative form for the ``cosmological  constant'' action,
\begin{equation}\label{eff}
S_{\rm eff}=-\frac{\Lambda}{8 \pi } \int d^4x ({}^{(4)}g)^{1/2}(g_{00})^{-2}~~~.
\end{equation}
This action is  Weyl scaling invariant, and in the absence of metric perturbations when $g_{00}(x)=1$ mimics standard $\Lambda$CDM cosmology, but is only 3-space generally coordinate invariant \cite{grav}.  I continued to assume that the part of the gravitational action involving metric derivatives is given by the 4-space generally coordinate invariant Einstein-Hilbert action (which is not Weyl scaling invariant). I pursued consequences of the action of Eq. \eqref{eff} in a number of papers from 2013 to the present. In \cite{horizon} written with Ramazano\u glu, we showed that as a result of the factor $g_{00}^{-2}$ in Eq. \eqref{eff}, the metric component $g_{00}$ never vanishes, leading to horizonless ``black'' holes, with an exterior geometry outside the nominal horizon very close to the standard Schwarzschild metric, but with a very different internal structure.  I followed this up with an expository paper on horizon structure for the Gravity Research Foundation Essay Competition  \cite{gravessay}, and with an analysis of metric perturbations \cite{rwpert} which shows that enough general coordinate invariance remains with the modified cosmological constant action to eliminate propagating scalar gravitational waves.  More recently, I looked at implications Eq. \eqref{eff} for the Hubble expansion rate \cite{hubble1}, \cite{hubble2}, for solar system relativity tests and lensing \cite{lens}, and with K. S.Virbhadra, for the photon sphere and black hole shadow radii \cite{shadow}.  Finally, at the invitation of K. K. Phua I wrote a mini-review for Modern Physics Letters A, which was under the editorship of Lars Brink, summarizing many of my papers on the possibility of a cosmological constant action that is not a quantum vacuum energy action \cite{mplA}.

\item{\bf Leaky ``black'' holes, horizonless ``dynamical  gravastar'' models, and astrophysical consequences}

My most recent work has focused on the possibility of horizonless black holes, as suggested initially by the modified cosmological action of Eq. \eqref{eff}.  An additional motivation comes from the underlying philosophy of trace dynamics, which asserts that at the fundamental level the universe is completely nonlocal, with all degrees of freedom in communication, and with locality only an approximation arising from thermodynamic averaging.  This world view conflicts with the prevailing idea that astrophysical black holes are true mathematical black holes, with interiors causally disconnected from exteriors; since each galaxy is believed now to contain a massive hole at its core, this would mean that the universe is divided into trillions of causally disconnected regions.  Based on the trace dynamics vantage point, I am deeply skeptical of this.

If black holes don't have event horizons \cite{fethi} or apparent horizons \cite{apparent} they can be ``leaky'', in the sense that interior matter can get out \cite{leaky}.  This suggests that leaky holes  can play a role in such astrophysical processes as young star formation near the central hole in our galaxy \cite{starform}, and  the formation of galaxies \cite{galaxy}, \cite{jwst}  themselves.  However, the cosmological constant is very small, suggesting that another mechanism may be at work if ``black'' hole leakage is to be of astrophysical importance.  This has motivated my most recent papers, which explore what I call ``dynamical gravastar'' models \cite{dynamical}, in which radii where structural changes occur come from solving the relativistic Tolman-Oppenheimer-Volkoff (TOV) equations with an assumed jump in the equation of state at high pressure, going from a relativistic matter equation of state in the exterior to one where pressure plus density sum to approximately zero in the interior.  Continuity properties of the TOV equations require the jump to be in the energy density, not in the pressure as generally assumed in the prior work reviewed in \cite{cardoso}.   Mathematica notebooks for a simplified gravastar model with zero cosmological constant are available for the user community at \cite{wolfram}, and give gravastars very closely approximating the exterior Schwarzschild geometry, with no horizons or trapped surfaces. Such models may evade \cite{evade} objections that have been raised to the idea of horizonless black hole mimickers, and have interesting mathematical properties that are the subject of my ongoing work.

\end{itemize}

\section{In appreciation of Lars Brink}

I met Lars Brink only very briefly a few times at the Institute for Advanced Study in Princeton, and at the Schwinger celebratory conference in Singapore.  But he touched on  my work nonetheless through his managing editorial roles at the World Scientific journals, where a number of the papers discussed above were published, with the benefit of constructive referee reports that led in some cases to substantial improvements.

The trace dynamics program and work flowing from it are a gamble: experiment has not yet spoken on the issues at hand.  But I deeply appreciate Lars's openness to such work, as exemplified in the following report I got on the invited mini-review that I wrote for Modern Physics Letters A: ``on behalf of: L Brink, Editor~~~Comments from the Editor and Reviewer:  This is a review article describing a program that the author has followed for quite some time. The idea is that dark energy is not a quantum vacuum energy, but rather arises from a Weyl scaling invariant nonderivative component of the gravitational action. The idea is shaking up our ideas of quantum gravity and cosmology, and the author is very skillfully navigating to answer all the questions that a skeptic reader might have.''  I was encouraged by this at the time and still am;  Lars Brink's open-mindedness and wisdom will be greatly missed.

\section{Acknowledgement}
This article is based  on a talk I gave in the ``Octonion Standard Model Unification''  (OSMU) internet lecture series, organized by Tejander Singh and Michael Wright.  The present writeup, as a contribution to the Lars Brink Memorial Volume,  benefited from the hospitality of the Aspen Center for Physics, which is supported by National Science Foundation Grant PHY-2210452.

\end{document}